\documentclass[12pt,a4paper]{article}
\usepackage[utf8]{inputenc}
\usepackage[T1]{fontenc}
\usepackage{amsmath}
\usepackage{amsfonts}
\usepackage{amssymb}
\usepackage{graphicx}
\usepackage{hyperref}
\usepackage[numbers,sort&compress]{natbib}
\usepackage{authblk}

\title{Challenge of nuclear transmutation in heavy-ion colliders}
\author[1,2]{I.A.~Pshenichnov}
\author[1,2]{S.D.~Savenkov}
\author[1,2]{A.O.~Svetlichnyi}
\affil[1]{Institute for Nuclear Research of the Russian Academy of Sciences, Moscow, Russia}
\affil[2]{Moscow Institute of Physics and Technology, Dolgoprudny, Russia}
\newcommand{\firstauthoremail}{pshenich@inr.ru}

\begin{document}

\date{}

\maketitle

\begin{center}
    \small
    Correspondence: \href{mailto:\firstauthoremail}{\firstauthoremail} (Igor Pshenichnov)
\end{center}

\begin{abstract}
We investigate the production of secondary nuclei in the hadronic fragmentation and electromagnetic dissociation (EMD) of $^{20}$Ne beams at the LHC and $^{124}$Xe beams at NICA. For light nuclei at LHC energies, our calculations show that hadronic interactions are the dominant channel for nuclear transmutation. This contrasts with the previously established dominance of EMD in $^{208}$Pb--$^{208}$Pb collisions. For $^{20}$Ne--$^{20}$Ne collisions at $\sqrt{s_{\rm NN}}=5.36$~TeV, we provide calculated cross sections and momentum distributions of produced nuclei such as $^4$He, $^{12}$C, $^{14}$N, and $^{16}$O. These results are essential for assessing potential contamination of the $^{20}$Ne--$^{20}$Ne event sample by collisions involving other nuclear species. Secondary nuclei will be also produced in the EMD of $^{124}$Xe beams at NICA, but they will not contaminate $^{124}$Xe--$^{124}$Xe data.
\end{abstract}

\vspace{0.5cm}

\maketitle

\section{Introduction}\label{intro}

The initial collision energy in heavy-ion colliders far exceeds the total binding energy of nucleons which confines them in interacting nuclei. One might therefore naively assume that such collisions result in the complete disintegration of beam nuclei into multiple individual nucleons and light nuclear fragments accompanied by numerous secondary hadrons. However, a study of 158A~GeV $^{208}$Pb projectiles interacting with heavy target nuclei at lower collision energy has demonstrated~\cite{Scheidenberger2004} that this is not the case for fragmentation of $^{208}$Pb in peripheral hadronic collisions and also for electromagnetic dissociation (EMD) of $^{208}$Pb. As follows from measurements and calculations~\cite{Scheidenberger2004}, heavy residual nuclei are produced frequently as spectator nuclei, and the production of Tl, Hg, Au, Pt, Ir, Os and other elements adjacent to lead in the periodic table is quite probable. This phenomenon has been extrapolated to the LHC energy, where the conversion of $^{208}$Pb beam nuclei to secondary nuclei of other elements with similar charge $Z$, mass $A$ or charge-to-mass ratio $Z/A$ were predicted in hadronic and EMD events~\cite{Pshenichnov2011a,Dmitrieva2023}. Similarly to proton-induced spallation reactions~\cite{David2015}, this can be considered as a phenomenon of nuclear transmutation in colliders. Recent ALICE measurements of the cross sections for the emission of one, two and three protons in EMD of $^{208}$Pb nuclei indicated their dominant transmutation to thallium, mercury and gold nuclei~\cite{Acharya2025}. 

In summer of 2025, $^{16}$O--$^{16}$O and $^{20}$Ne--$^{20}$Ne collisions at $\sqrt{s_{\rm NN}}=5.36$~TeV were studied at the LHC for the first time. In the present work, the production of secondary spectator nuclei in hadronic $^{20}$Ne--$^{20}$Ne collisions at the LHC was modeled by means of the Abrasion-Ablation Monte Carlo for Colliders model (AAMCC) with Minimum Spanning Tree (MST) clustering ~\cite{Kozyrev2022,Nepeivoda2022}. These results supplement the \mbox{AAMCC-MST} predictions given earlier~\cite{Svetlichnyi2023} for hadronic $^{16}$O--$^{16}$O collisions at the LHC. Finally, the EMD of $^{124}$Xe in ultraperipheral $^{124}$Xe--$^{124}$Xe collisions at NICA accelerator complex was also modeled to assess the transmutation of beam nuclei at much lower collision energy which will be available at NICA.

\section{Modeling hadronic fragmentation and EMD of relativistic nuclei}

In this work, nucleus-nucleus collisions at the LHC were simulated by means of the AAMCC-MST model~\cite{Kozyrev2022,Nepeivoda2022,Svetlichnyi2023}. This model is focused on the production of spectator nucleons and fragments. It is based on the combination of (1) Glauber Monte Carlo model~\cite{Loizides2018} to calculate the total volume of spectator matter (prefragments) in each collision; (2) MST-clustering algorithm to model pre-equilibrium decays of prefragments;  and (3) statistical models to simulate the decays of excited prefragments and clusters~\cite{Kozyrev2022,Nepeivoda2022,Svetlichnyi2023}. 

The RELDIS model~\cite{Pshenichnov2011} is used in this work to simulate the EMD or relativististic nuclei. It is based on a photonuclear reaction model~\cite{Ilinov1996, Pshenichnov2005}, which simulates the interaction of a photon with a nucleus as a four-step process. The modeling begins with (1) an intranuclear cascade of particles induced by the photon inside the nucleus, followed by (2) pre-equilibrium emission of nucleons, followed in some events by (3) nucleon coalescence, and finally (4) the excited residual nucleus evaporates nucleons or undergoes fission.

The relation between hadronic fragmentation and EMD is different in collisions of light and heavy nuclei. For the initial evaluation of this relation, the total hadronic reaction cross sections and the total EMD cross sections can be considered. These cross sections are listed in Table~\ref{tab1:total} for  $^{16}$O--$^{16}$O, $^{20}$Ne--$^{20}$Ne and $^{208}$Pb--$^{208}$Pb collisions at the LHC and $^{124}$Xe--$^{124}$Xe collisions at NICA. 
\begin{table}[htb!]
\renewcommand{\arraystretch}{1.25}
\renewcommand{\tabcolsep}{3pt}
\caption{Total hadronic and EMD cross sections for $^{16}$O--$^{16}$O,    $^{20}$Ne--$^{20}$Ne and $^{208}$Pb--$^{208}$Pb  collisions at the LHC and  $^{124}$Xe--$^{124}$Xe collisions at NICA}
\begin{tabular}{|c|c|c|c|}
\hline
colliding & $\sqrt{s_{\rm NN}}$ & \multicolumn{2}{|c|}{total cross sections (b)}  \\
\cline{3-4}
nuclei & TeV &  hadronic & EMD \\
\hline
$^{16}$O--$^{16}$O & 5.36  & 1.3  & 0.12        \\
\hline
$^{20}$Ne--$^{20}$Ne & 5.36  & 1.7  & 0.3   \\
\hline
$^{208}$Pb--$^{208}$Pb &  5.36 & 7.6 & 206.5 \\
\hline
$^{124}$Xe--$^{124}$Xe &  0.0095 & 6 &  5.3 \\
\hline
\end{tabular}
\label{tab1:total}
\end{table}

The total hadronic and EMD cross sections were calculated, respectively, with Glauber Monte Carlo and RELDIS models. There are large uncertainties of the total cross sections for the photoabsorption  on $^{16}$O and $^{20}$Ne target nuclei measured in experiments with real photons. These cross sections were used in RELDIS modeling as an input. Therefore, the results for the total EMD cross sections for $^{16}$O and $^{20}$Ne nuclei can be less reliable in comparison to the EMD cross sections for $^{208}$Pb already validated by measurements at the LHC~\cite{Acharya2023}. Nevertheless, as follows from Table~\ref{tab1:total}, the dominance of the total hadronic cross sections over the total EMD cross sections in $^{16}$O--$^{16}$O and  $^{20}$Ne--$^{20}$Ne collisions is obvious. In contrast, the production of heavy secondary nuclei in hadronic $^{208}$Pb--$^{208}$Pb collisions is of minor importance compared to EMD~\cite{Acharya2023,Acharya2025}. Heavy spectator fragments can be produced by removal of a very few nucleons from colliding nuclei mostly in very peripheral collisions, but their cross sections are at the level of few percent of the total hadronic cross section.   

\section{$^{208}$Pb--$^{208}$Pb Collisions at the LHC}
Recent ALICE measurements~\cite{Acharya2025} have confirmed the production of heavy residual nuclei, Tl, Hg and Au, in the EMD of $^{208}$Pb at the LHC~\cite{Pshenichnov2011a,Dmitrieva2023}. The total yield of gold nuclei has been explicitly evaluated in Ref.~\cite{Acharya2025}. This attracted a lot of attention and has been referred to as alchemy at the LHC~\cite{Gybney2025,Afshar2025}. Because of the very large cross sections to produce $^{206}$Pb and $^{207}$Pb in the EMD of $^{208}$Pb~\cite{Acharya2023}, these isotopes of lead pose a special risk to the LHC operation and limit the beam life time and collider luminosity~\cite{Bruce2009}. 

As estimated with RELDIS~\cite{Pshenichnov2011a}, among the secondary nuclei produced in the EMD of $^{208}$Pb, there are quite a lot of nuclei with a charge-to-mass ratio close to that of $^{208}$Pb. Since this ratio determines the trajectories of the nuclei as they pass through the magnetic field of the collider, such nuclei can propagate large distances from the interaction point, but will eventually be intercepted by the LHC collimators. However, none of the secondary nuclei have a charge-to-mass ratio exactly equal to that of $^{208}$Pb. Consequently, none of the EMD products could circulate in the collider ring together with  beam nuclei, collide with nuclei of the  counter-rotating beam at the interaction points and thus contaminate the data.

\section{$^{20}$Ne--$^{20}$Ne collisions at the LHC}
\begin{figure}[htb]
\centering
    \includegraphics[width=1.05\linewidth]{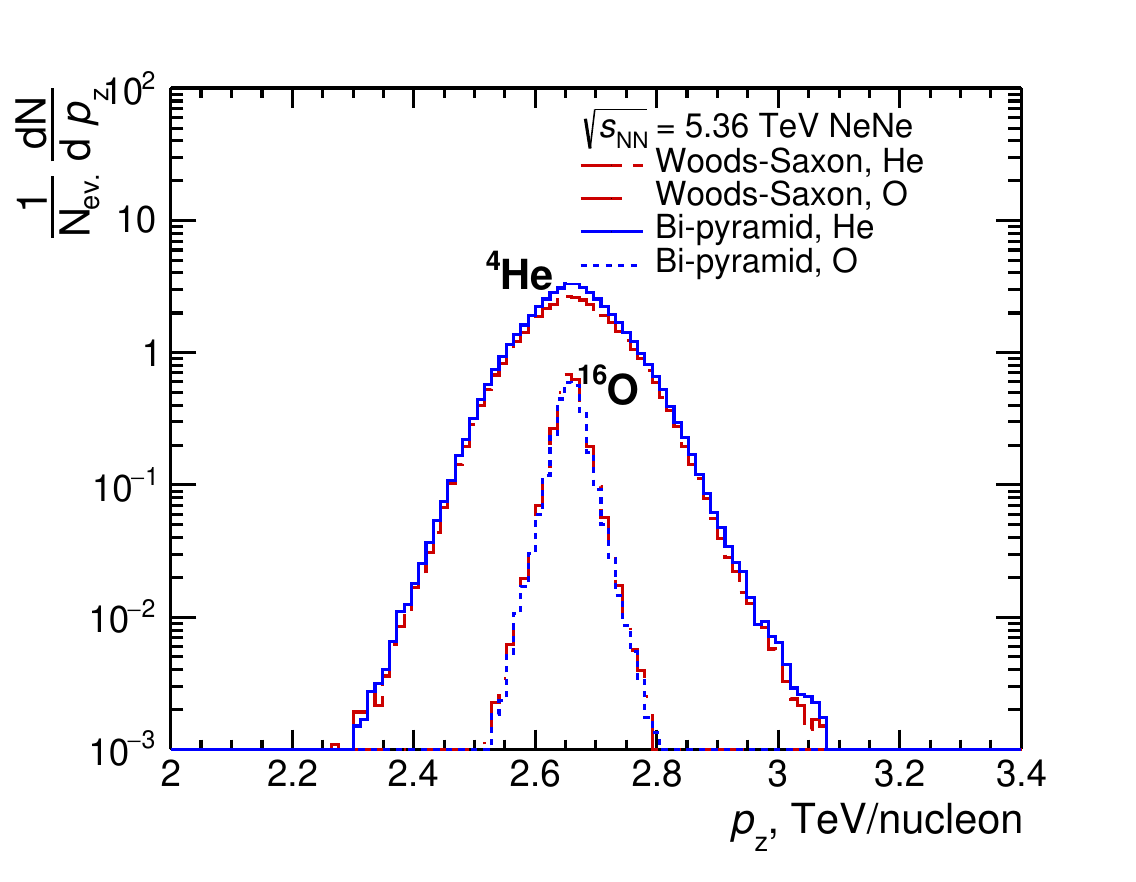}
\caption{\label{fig:pz_o_he} Longitudinal momentum distributions of secondary $^4$He and $^{16}$O nuclei produced in hadronic fragmentation in $^{20}$Ne--$^{20}$Ne collisions at $\sqrt{s_{\rm NN}}=5.36$~TeV.}
\end{figure}

According to Table~\ref{tab1:total}, it is important to study first hadronic $^{20}$Ne--$^{20}$Ne collisions as a source of forward-going secondary nuclei at the LHC. Nuclei with the charge-to-mass ratio $Z/A=1/2$ equal to that of $^{20}$Ne are of primary interest, as they can potentially co-circulate with $^{20}$Ne in the collider.  The fragmentation of $^{20}$Ne nuclei was modeled with AAMCC-MST by considering two options for the nuclear matter density distribution in $^{20}$Ne: (1) a deformed Woods-Saxon distribution; and (2) assuming an alpha-clustered configuration in $^{20}$Ne represented by a bi-pyramid~\cite{Bijker2021}. The average inclusive cross sections of production of $^4$He, $^{12}$C, $^{14}$N and $^{16}$O calculated with these two options are equal to $1.04\pm 0.11$~b, $0.11\pm 0.01$~b, $0.17\pm 0.05$~b and $0.08\pm 0.02$~b, respectively. The longitudinal momentum distributions of $^4$He and $^{16}$O  are presented in Fig.~\ref{fig:pz_o_he}. 

\section{$^{124}$Xe--$^{124}$Xe Collisions at NICA}
As follows from Table~\ref{tab1:total}, the total hadronic and EMD cross sections are comparable for $^{124}$Xe--$^{124}$Xe collisions at NICA. However, in these collisions, as in the case of $^{208}$Pb--$^{208}$Pb collisions, the specific channels leading to nuclei with $Z$ and $A$ close to $^{124}$Xe have partial cross sections that are only a minor fraction of the total hadronic cross section. This motivates to study first the EMD of $^{124}$Xe nuclei at NICA.
\begin{figure}[htb]
\centering
    \includegraphics[width=1.05\linewidth]{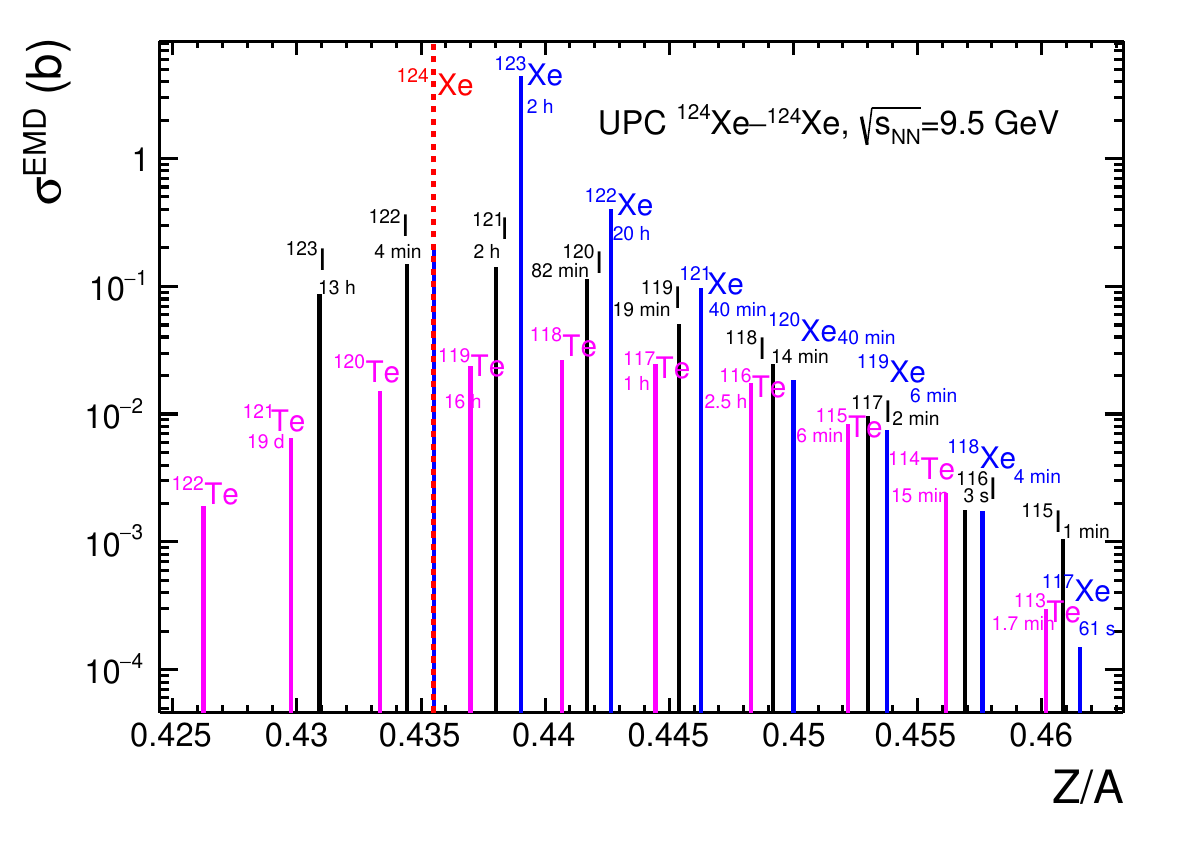}
\caption{\label{fig:XeXe124_95} Cross sections to produce secondary nuclei with a given charge-to-mass ratio $Z/A$ in EMD of $^{124}$Xe nuclei in their ultraperipheral collisions at  $\sqrt{s_{\rm NN}}=9.5$~GeV at NICA. The $Z/A$ ratio for $^{124}$Xe beam nuclei is marked by a dashed line}
\end{figure}

The cross sections to produce secondary nuclei with a given charge-to-mass number ratio $Z/A$ in EMD of $^{124}$Xe at  $\sqrt{s_{\rm NN}}=9.5$~GeV at NICA are presented in Fig.~\ref{fig:XeXe124_95}. Similarly to the case of $^{208}$Pb--$^{208}$Pb, none of the secondary nuclei are characterized by $Z/A$ exactly equal to that of beam nuclei. This means that the contamination of $^{124}$Xe beams is out of the question at NICA, but a possible impact of $^{122,123}$Xe and $^{120,121,122}$I nuclei on collider components requires further studies.

\section*{Conclusion}
In $^{124}$Xe--$^{124}$Xe 
collisions at NICA, the EMD of $^{124}$Xe beam nuclei is the dominant source of  secondary nuclei with charge $Z$ and mass
$A$ similar to $^{124}$Xe. However, none of these secondary nuclei possess the same charge-to-mass ratio $Z/A$ as $^{124}$Xe. Since NICA will be tuned to collide ions with the specific $Z/A$ of $^{124}$Xe, this prevents other nuclei from co-circulating with primary beams and contaminating pure $^{124}$Xe--$^{124}$Xe collision events with collisions involving other nuclear species.    

In contrast, it is expected that certain nuclei with $Z/A=1/2$ equal to $Z/A$ of $^{20}$Ne are produced in hadronic $^{20}$Ne--$^{20}$Ne collisions at the LHC. These nuclei can co-circulate with $^{20}$Ne nuclei and potentially contaminate $^{20}$Ne--$^{20}$Ne data with $^{4}$He--$^{20}$Ne, $^{12}$C--$^{20}$Ne, $^{14}$N--$^{20}$Ne, $^{16}$O--$^{20}$Ne events. 

\bibliographystyle{elsarticle-num}

\bibliography{Transmutation-NICA-LHC}

\end{document}